\documentclass[12pt]{article}

\usepackage{jheppub} 
\pdfoutput=1
\usepackage{xcolor}
\usepackage{subcaption}
\usepackage[utf8]{inputenc}
\usepackage{graphicx}
\usepackage{verbatim}
\usepackage{tensor}
\usepackage{setspace}
\usepackage{amsmath, amssymb, amsthm, float, graphicx, amsfonts, dsfont}
\numberwithin{equation}{section}

\newcommand{\be}{\begin{equation}}
\newcommand{\ee}{\end{equation}}

\def\G{\Gamma}

\def\a{\alpha}

\def\tt{\tilde{t}}

\def\b{\beta}
\def\d{\delta}

\def\f{\phi}

\def\D{\Delta}

\newcommand{\bea}{\begin{eqnarray}}
\newcommand{\eea}{\end{eqnarray}}

\renewcommand{\a}{\alpha}
\renewcommand{\b}{\beta}

\renewcommand{\d}{\delta}
\newcommand{\dsl}{\pa \kern-0.5em /}

\newcommand{\pa}{\partial}

\newcommand{\nn}{\nonumber\\}

\renewcommand{\eqref}[1]{eq.(\ref{#1})}

\abstract{We consider holographic CFTs and study their large $N$ expansion. We use Polyakov-Mellin bootstrap to extract the CFT data of all operators, including scalars, till $O(1/N^4)$. We add a contact term in Mellin space, which corresponds to an effective $\phi^4$ theory in AdS and leads to anomalous dimensions for scalars at $O(1/N^2)$. Using this we fix $O(1/N^4)$ anomalous dimensions for double trace operators finding perfect agreement with \cite{loopal} (for $\Delta_{\phi}=2$). Our approach generalizes this to any dimensions and any value of conformal dimensions of external scalar field. In the second part of the paper, we compute the loop amplitude in AdS which corresponds to non-planar correlators of in CFT. More precisely, using CFT data at $O(1/N^4)$ we fix the AdS bubble diagram and the triangle diagram for the general case.}

\title{\bf Polyakov-Mellin Bootstrap for AdS loops}
\date{}

\author{\!\!\!\!  Kausik Ghosh\footnote{kau.rock91@gmail.com}\\ ~~~~\\
\it Centre for High Energy Physics,
\it Indian Institute of Science,\\ \it C.V. Raman Avenue, Bangalore 560012, India. }

\begin{document}

\maketitle

\section{Introduction}
We have seen a tremendous progress in constraining conformal field theories in space-time dimensions greater than two since the work of \cite{Rattazzi:2008pe}. The CFT data for operators with low conformal dimension has been calculated both numerically \cite{3dising,mostprecise,rychkovrev} and analytically \cite{rychkovtan,dsd,Komargodski:2012ek,Alday:2016njk}. An alternative approach, now called Polyakov-Mellin (PM) Bootstrap motivated by the work of Polyakov \cite{polya} (also see \cite{KSAS} for further extension) was proposed in \cite{RGAKKSAS,PDAKAS}. Usually bootstrap equation arises by matching different OPE expansions of correlation functions and demanding crossing symmetry. In contrast, one expands the correlation function in a manifestly crossing symmetric basis in PM bootstrap. Then demanding consistency with the physical OPE expansion gives rise to consistency equations. This problem was formulated in Mellin space which makes the pole structure transparent and also properties of certain orthogonal polynomials called continuous Hahn make the computation much simpler. Since the crossing symmetric basis is formed by the sum of witten exchange diagrams and contact terms, this method has limitations unless the contact terms are fixed systematically. This issue was first pointed out in \cite{PDKGAS} and in \cite{pmb} it was systematically explored in a couple of examples.\\

In this paper we will focus on holographic CFTs aka CFTs with large number of degrees of freedom $(N)$. We want to study large N expansion of the four point function using  Polyakov-Mellin bootstrap. In \cite{pmb}, the authors considered any local theory involving scalars in AdS and calculated the correction to conformal dimensions at $O(1/N^2)$, $\gamma^{(1)}_{n,\ell}$ using this formalism and recovered the results of \cite{kss}. After briefly reviewing it, we will calculate the $O(1/N^4)$ corrections to conformal dimensions in this model. Although at $O(1/N^2)$ only finite number of operators with non-zero anomalous dimension present depending on the number of contact diagrams are present in the bulk, at $O(1/N^4)$ we will see that $\gamma^{(2)}_{n,\ell}$ is non-zero for all spins. Therefore we will solve it using Polyakov-Mellin bootstrap techniques and will find  closed form expressions for coefficients appearing in  the large spin limit of $\gamma^{(2)}_{n,\ell}$. We will see that our results match perfectly with the existing results \cite{loopal} derived using usual bootstrap techniques. As a bonus our final expressions are valid for any general $\Delta_{\phi}$ and space-time dimension. In \cite{loopal} this was solved for integer values of $\Delta_{\phi}$. Also our formula trivially extends to any space-time dimensions whereas it is not at all straightforward in usual bootstrap techniques because of unavailability of closed form expressions for conformal blocks in odd dimensions.\\

In the second half of the paper we will use the AdS/CFT correspondence to fix a few AdS loop diagrams following \cite{loopal}. Basically the large N expansions of correlator gets mapped to perturbative expansions in AdS space. There are quite a few papers on loops \cite{costaloop,kaplanloop1,kaplanloop2,loop3,loop4,loop5,loop6,Giombi:2017hpr} but evaluation of integrations in AdS pose technical complications and the methods of this paper for reconstructing loop from CFT data might be very useful to make further progress. Since we have a closed form expression for $\gamma^{(2)}_{0,\ell}$, in principle, we can fix the loop amplitudes for AdS scalar theories for any general case quite easily.

\section{Polyakov Mellin Bootstrap}
In this section we give a brief overview of working rules of Polyakov Mellin bootstrap. We focus on four point functions of scalar fields $\phi(x)$ in this paper. Conformal symmetry fix the structure of four point function as follows,

\begin{equation} \label{sch} 
\langle\underbrace{\phi(x_1)\phi(x_2)}\underbrace{\phi(x_3)\phi(x_4)}\rangle=\frac{1}{x^{2\Delta_{\phi}}_{12}x^{2\Delta_{\phi}}_{34}}\sum_{\Delta,\ell}C_{\Delta,\ell}\,\, g_{\Delta,\ell}(u,v)
\end{equation}
Here we used the OPE of $\phi(x_1)\phi(x_2)$ and $\phi(x_3)\phi(x_4)$ which is expressed in terms of $s-$channel conformal blocks. We could have used the OPE of $\phi(x_1)\phi(x_4)$ and $\phi(x_2)\phi(x_3)$ and express it in terms of $t-$channel conformal blocks.
\begin{equation}
\langle\underbrace{\phi(x_1)\overbrace{\phi(x_2)\phi(x_3)}\phi(x_4)}\rangle=\frac{1}{x^{2\Delta_{\phi}}_{14}x^{2\Delta_{\phi}}_{23}}\sum_{\Delta,\ell}C_{\Delta,\ell} \,\, g_{\Delta,\ell}(v,u)
\end{equation}
 Since we are expanding the same four point function the two conformal block decomposition should be same but this is not manifest. Demanding this equivalence give us the usual bootstrap condition (crossing equation),
\begin{equation}
\sum_{\Delta,\ell}C_{\Delta,\ell} g_{\Delta,\ell}(u,v)=\left(\frac{u}{v}\right)^{\Delta_{\phi}} \sum_{\Delta,\ell}C_{\Delta,\ell} g_{\Delta,\ell}(v,u)
\end{equation}
Now one can think of a block decomposition which is crossing symmetric on its own (e.g.. manifestly symmetric under $u\leftrightarrow v $). It is a legitimate thing to do to add all three channel ($s$,$t$ and $u$) and expand four point function in that basis which would also be consistent with OPE. But we would not be able to use this directly to find CFT data $(\Delta,C_{\Delta,\ell})$. Instead one needs to add $s$,$t$, and $u$ channel witten blocks which is a crossing symmetric basis on its own. But then since Witten blocks are not guaranteed to be consistent with OPE. One will find contributions from double trace operators (with exact dimension $2\Delta_{\phi}+2n$) which are absent in the OPE. Cancellation of this contribution constraint the spectrum of CFT. We will work with this basis in Mellin space. The Mellin transformation of four point function is defined as\footnote{we are using the notation,$[ds]=\frac{ds}{2\pi i}$ and $[dt]=\frac{dt}{2\pi i}$ },
\begin{equation} \label{mellin}
\langle\phi(x_1)\phi(x_2)\phi(x_3)\phi(x_4)\rangle=\frac{1}{x_{12}^{2\Delta_{\phi}}x_{34}^{2\Delta_{\phi}}}\int [ds][dt]  u^s v^t \rho_{\Delta_{\phi}}(s,t) M(s,t)
\end{equation} 
where, the Mellin amplitude $M(s,t)$ is defined as\footnote{$c_{\Delta,\ell}=C_{\Delta,\ell}\mathcal{N}_{\Delta,\ell}$, where $C_{\Delta,\ell}$ is the usual OPE coefficient and $\mathcal{N}_{\Delta,\ell}$ is defined in Appendix B.},
\begin{equation}
M(s,t)=\sum_{\Delta,\ell}c_{\Delta,\ell}\left(W_{\Delta,\ell}^{s}(s,t)+W_{\Delta,\ell}^t(s,t)+W_{\Delta,\ell}^u(s,t)\right),
\end{equation}
these are $s$,$t$ and $u$ exchange Witten blocks, and the measure is given by,
\begin{equation}
\rho_{\Delta_{\phi}}(s,t)=\Gamma^2(\Delta_{\phi}-s) \Gamma^2(-t)\Gamma^2(s+t).
\end{equation}

\subsection{PM consistency conditions}

Recently in \cite{pmb} the spectral integrals were explicitly evaluated. The $s-$channel block is given by,
\begin{equation}
W_{\Delta,\ell}^{s}(s,t)=- 2 \frac{f_{p}^{\ell}(s,t)\Gamma(\frac{\Delta+\ell}{2}+\Delta_{\phi}-h)^2}{(\ell+2s-\Delta)\Gamma(\Delta-h+1)}\,{}_3F_2\bigg[\begin{matrix} \frac{\Delta-\ell} {2}-s,\, 1+\frac{\Delta-\ell} {2}-\Delta_{\phi},\,1+\frac{\Delta-\ell} {2}-\Delta_{\phi}\\
	\ \ 1+\frac{\Delta-\ell} {2}-s \ \ , \ \ \ \ \ \  \Delta-h+1
\end{matrix};1\bigg]\,.
\end{equation}
In our notation,$h=\frac{d}{2}$ \footnote{$f_{p}^{\ell}(s,t)=\frac{P^{(s)}_{\Delta-h,\ell}(s,t)}{(\Delta-1)_{\ell}(2h-\Delta-1)_{\ell}}$. $P^{(s)}_{\Delta-h,\ell}(s,t)$ is the Mack polynomial and we follow the same convention as \cite{pmb}.}.The $t-$channel block is given by replacing $s\rightarrow t+\Delta_{\phi}$ and $t\rightarrow s-\Delta_{\phi}$ and $u-$channel is given by replacing $s\rightarrow\Delta_{\phi}-s-t$ and $t\rightarrow t$ in $s-$channel respectively. It can be seen from equation (\ref{mellin}) that we will get powers like $u^{\Delta_{\phi+r}}\log u$, $r$ being any non-negative number, which comes from double poles $\Gamma^2(\Delta_{\phi}-s)$ in from the measure. Similarly we will also get $u^{\Delta_{\phi+r}}$ for single pole contribution from the same measure.  These are absent in the physical OPE of $\phi \times \phi$. So we will get bootstrap condition which needs to satisfy for each r separately, from the cancellation of residue at double poles in s we get,
\begin{equation} \label{doubleeq}
\sum_{\Delta,\ell}c_{\Delta,\ell} \left(W^{s}_{\Delta,\ell}(\Delta_{\phi}+r,t)+W^{t}_{\Delta,\ell}(\Delta_{\phi}+r,t)+W^{u}_{\Delta,\ell}(\Delta_{\phi}+r,t)\right)=0
\end{equation}
and from the cancellation of residue at single poles in s we get,
\begin{equation} \label{singlepole}
\sum_{\Delta,\ell}c_{\Delta,\ell} \left(W^{s'}_{\Delta,\ell}(\Delta_{\phi}+r,t)+W^{t'}_{\Delta,\ell}(\Delta_{\phi}+r,t)+W^{u'}_{\Delta,\ell}(\Delta_{\phi}+r,t)\right)+W^{'}_{0,0} =0
\end{equation}
where $'$ indicates derivative with respect to $s$ and $W^{'}_{0,0}$ stands for the single pole contribution from identity exchange which doesn't contribute to the residue at double pole. Though in this paper we will only focus in solving the equation (\ref{doubleeq}) which will give us anomalous dimensions which are important and sufficient to reconstruct the loops in AdS, one can solve the (\ref{singlepole}) equations analogously to find corrections to OPE coefficients.

 Now it turns out that it is useful to decompose the blocks in continuous Hahn Polynomial in $t$, so that we can write,
\begin{equation}
W_{\Delta,\ell}^{s}(s,t)= \sum_{\ell'}q^{s}_{\Delta,\ell'|\ell}(s)Q^{2s+\ell'}_{\ell',0}(t).
\end{equation}
Similarly,
\begin{equation}
\begin{split}
& W_{\Delta,\ell}^{t}(s,t)= \sum_{\ell'}q^{t}_{\Delta,\ell'|\ell}(s)Q^{2s+\ell'}_{\ell',0}(t),\\
& W_{\Delta,\ell}^{u}(s,t)= \sum_{\ell'}q^{u}_{\Delta,\ell'|\ell}(s)Q^{2s+\ell'}_{\ell',0}(t).
\end{split}
\end{equation}
These continuous Hahn polynomial satisfy orthogonality relation in $\ell'$ given in Appendix A. Therefore we get the following bootstrap condition for each $\ell'$,
\begin{equation}
\bigg(\sum_{\Delta,\ell}c_{\Delta,\ell}q^{s}_{\Delta,\ell'|\ell}(s)+\sum_{\Delta,\ell}c_{\Delta,\ell} q^{t}_{\Delta,\ell'|\ell}(s)+ \sum_{\Delta,\ell}c_{\Delta,\ell} q^{u}_{\Delta,\ell'|\ell}(s)\bigg)|_{s=\Delta_{\phi}+r} =0.
\end{equation}
and\footnote{$q_{0,0}(s)$ represents the identity contribution after decomposition in continuous Hahn basis.}
\begin{equation}
\bigg(\sum_{\Delta,\ell}c_{\Delta,\ell}q^{s'}_{\Delta,\ell'|\ell}(s)+\sum_{\Delta,\ell}c_{\Delta,\ell} q^{t'}_{\Delta,\ell'|\ell}(s)+ \sum_{\Delta,\ell}c_{\Delta,\ell} q^{u'}_{\Delta,\ell'|\ell}(s)+q^{'}_{0,0}(s)\bigg)|_{s=\Delta_{\phi}+r} =0.
\end{equation}
Since for identical scalars we have the symmetry,
\begin{equation}
W_{\Delta,\ell}^{t}(s,t)=W_{\Delta,\ell}^{u}(s,-s-t),
\end{equation}
which is also the symmetry of continuous Hahn polynomial, we conclude that,
\begin{equation}
q^{t}_{\Delta,\ell'|\ell}(s)=q^{u}_{\Delta,\ell'|\ell}(s).
\end{equation}
So we get our final form of the consistency equation is ,
\begin{equation} \label{mellineq}
\bigg(\sum_{\Delta,\ell}c_{\Delta,\ell}q^{s}_{\Delta,\ell'|\ell}(s)+2\sum_{\Delta,\ell}c_{\Delta,\ell} q^{t}_{\Delta,\ell'|\ell}(s)\bigg)|_{s=\Delta_{\phi}+r} =0,
\end{equation}
and
\begin{equation} \label{mellineqs}
\bigg(\sum_{\Delta,\ell}c_{\Delta,\ell}q^{s'}_{\Delta,\ell'|\ell}(s)+2\sum_{\Delta,\ell}c_{\Delta,\ell} q^{t'}_{\Delta,\ell'|\ell}(s)+q^{'}_{0,0}(s)\bigg)|_{s=\Delta_{\phi}+r} =0.
\end{equation}
The compact expressions for $q^s_{\Delta,\ell'|\ell}$ and $q^t_{\Delta,\ell'|\ell}$ were derived in \cite{pmb} which we quote below,
\begin{equation}
\begin{split}
q^s_{\Delta,\ell'|\ell}=& \sum_{m,n}\mu^{(\ell)}_{m,n} \left(\frac{\Delta-\ell}{2}-s\right)_m \chi^{(n)}_{\ell'}(s)\frac{\Gamma^2(\frac{\Delta+\ell}{2}+\Delta_{\phi}-h)}{\left(\frac{\Delta-\ell}{2}-s\right)\Gamma(\Delta-h+1)}\\
& \,_3F_2[\{\frac{\Delta-\ell}{2}-s,1+\frac{\Delta-\ell}{2}-\Delta_{\phi},1+\frac{\Delta-\ell}{2}-\Delta_{\phi}\},\{1+\frac{\Delta-\ell}{2}-s,\Delta-h+1\},1].
\end{split}
\end{equation}
\begin{equation} \label{tchexp}
\begin{split}
q^t_{\Delta,\ell'|\ell}=&\sum_{m,n}(-1)^{\ell'+m} 2^{-\ell'} \mu_{m,n}^{\ell}\left(\Delta_{\phi}-s\right)_n \left(a_{\ell}\right)_m^2 \Gamma(2s+2\ell')\Gamma^2(d) \Gamma(\frac{a}{2})\Gamma(a+1)\Gamma^2(1+a-f-b)\\
& \,_7F_6 \left(\begin{matrix} a, & 1+\frac{1}{2}+a, & b, & c, & d, & e, & f \\ \frac{1}{2}a, & 1+a-b, & 1+a-c, & 1+a-d, & 1+a-e, & 1+ a-f 
\end{matrix} ;1\right).
\end{split}
\end{equation}
The expressions of $\chi^{(n)}_{\ell}(s)$, and a,b,c etc. are given in Appendix B.
\section{ PM bootstrap at Large N }

Now we turn our attention to bootstrap holographic theories by adding appropriate contact terms. Since we are assuming that we are working in large N limit, therefore a large class of theories are there which will have only identity and double field operators in the leading order. More precisely, In the leading order in large N limit identity and  only double field operators appear with its classical twist and mean field OPE coefficient as follows,

\begin{equation} \label{meanfield}
\begin{split}
& \tau^{(0)}_{n,\ell}=2\Delta_{\phi}+2n\\
& C^{(0)}_{n,\ell}=\frac{\left((-1)^{\ell }+1\right) \left(\left(-\frac{d}{2}+\Delta \phi +1\right)_n\right){}^2 \left((\Delta \phi )_{n+\ell }\right){}^2}{n! \ell ! (-d+n+2 \Delta \phi +1)_n \left(\frac{d}{2}+\ell \right)_n (2 n+\ell +2 \Delta \phi -1)_{\ell } \left(-\frac{d}{2}+n+\ell +2 \Delta \phi \right)_n}
\end{split}
\end{equation}
As we go to next sub-leading order these operators start receiving quantum correction as anomalous dimensions and also the OPE coefficients get corrected,
\begin{equation}\label{correction}
\begin{split}
&\tau_{n,\ell}=2\Delta_{\phi}+2n+\frac{\gamma^{(1)}_{n,\ell}}{N^2}+\frac{\gamma^{(2)}_{n,\ell}}{N^4}+....\\
&  C_{n,\ell}= C^{0}_{n,\ell}+\frac{C^{(1)}_{n,\ell}}{N^2}+\frac{C^{(2)}_{n,\ell}}{N^4}+....
\end{split}
\end{equation}
 In the next two subsections we describe in detail the calculation of these corrections.

\subsection{Analysis at $O(1/N^2)$}

First we consider $\phi^4$ theory in the bulk. Therefore we add this contact term to our basis. The Mellin amplitude corresponding to this contact term is a constant. Discussion of more general cases where the contact diagram is a polynomial in s and t and not just a constant is relegated to appendix \ref{contact}. The $\phi^4$ term is decomposed  in continuous Hahn basis and since it is a constant\footnote{if it is a polynomial of degree $n$ in mellin variable $t$ then that contact term will contribute from $\ell'=0$ to $\ell'=n$.} it will contribute to only $\ell'=0$ equation,

\begin{equation}
\lambda=\sum_{\ell'} \lambda \delta_{\ell',0} Q^{2s+\ell'}_{\ell',0}\,.
\end{equation}

 So our constraint equation (\ref{mellineq}) at $s=\Delta_{\phi}$ and $\ell'=0$ becomes,
\begin{equation}
\sum_{\Delta,\ell}c_{\Delta,\ell}\Big(q^{s}_{\Delta,0|\ell}+2q^{t}_{\Delta,0|\ell}\Big)+\lambda=0
\end{equation}
where $\lambda$ is coming from contact term and  we can use this equation to fix $\lambda$ which is,
\begin{equation} \label{fixc}
\lambda=-\sum_{\Delta,\ell}c_{\Delta,\ell}\Big(q^{s}_{\Delta,0|\ell}+2q^{t}_{\Delta,0|\ell}\Big)
\end{equation}
By plugging in the data given in equation (\ref{correction}) and expanding it to $O(1/N^2)$ we find,
\begin{equation} \label{s1}
\lambda=-\frac{2^{2 \Delta_{\phi} -1}  \Gamma \left(\Delta_{\phi} +\frac{1}{2}\right)}{\sqrt{\pi } \Gamma (\Delta_{\phi} )^3}\frac{\gamma^{(1)}_{0,0}}{N^2}
\end{equation}
The term on the right came from just $s-$channel. Expanding the $t$- channel we found that it starts at $O(1/N^4)$ because of the suppression for double trace operators coming from $\mathcal{N}_{\Delta,\ell}$. This procedure of fixing contact term was first pointed out in \cite{pmb}.\\
The operators $\Delta_{0,0}$ and $\Delta_{1,0}$ contribute at leading order when we evaluate $s-$channel at $s=\Delta_{\phi}+1$ and here also we have a term coming from contact term and again crossed channel starts contributing from $O(1/N^4)$. Taking all the contributions till $O(1/N^2)$ we have ,
\begin{equation} \label{s2}
\frac{2^{2 \Delta_{\phi} -1}  \Gamma \left(\Delta_{\phi} +\frac{1}{2}\right)}{\sqrt{\pi } \Gamma ^3(\Delta_{\phi} ) (2 \Delta_{\phi} -h+1)}\frac{\gamma^{(1)}_{0,0}}{N^2}+\frac{2^{2 \Delta_{\phi} +1} \Delta_{\phi} ^2  (2h-2 (\Delta_{\phi} +1)) \Gamma \left(\Delta_{\phi} +\frac{3}{2}\right)}{\sqrt{\pi } 2h (2h-4 \Delta_{\phi} -2) \Gamma^3 (\Delta_{\phi} +1)}\frac{\gamma^{(1)}_{1,0}}{N^2}+\lambda=0~\,.
\end{equation}
Now we replace $\lambda$ using (\ref{fixc}) to find,
\begin{equation} \label{g1}
\gamma^{(1)}_{1,0}=\frac{\Delta_{\phi}  h (h-2 \Delta_{\phi} )}{(2 \Delta_{\phi} +1) (-\Delta_{\phi} +h-1)}\gamma^{(1)}_{0,0}.
\end{equation} 
We also include the equations evaluated at $s=\Delta_{\phi}+2$, where  at $O(1/N^2)$, the operators with dimensions $\Delta_{0,0}$,$\Delta_{1,0}$ and $\Delta_{2,0}$ contribute in $s-$channel,

\begin{equation} \label{s3}
\begin{split}
& \frac{4^{\Delta_{\phi} }  \Gamma \left(\Delta_{\phi} +\frac{1}{2}\right)}{\sqrt{\pi } \Gamma^3 (\Delta_{\phi} ) (-2 \Delta_{\phi} +h-1) (h-2 (\Delta_{\phi} +1))}\frac{\gamma^{1}_{0,0}}{N^2}+\frac{2^{2 \Delta_{\phi} +3} \Delta_{\phi} ^2  (-2h+2 \Delta_{\phi} +2) \Gamma \left(\Delta_{\phi} +\frac{3}{2}\right)}{\sqrt{\pi } 2h (2h-4 \Delta_{\phi} -2) \Gamma ^3(\Delta_{\phi} +1) (-2 \Delta_{\phi} +h-3)}\\
&\times \frac{\gamma^{1}_{1,0}}{N^2} -\frac{4^{\Delta_{\phi} +2} \Delta_{\phi} ^2 (\Delta_{\phi} +1)^2  (-2h+2 \Delta_{\phi} +4) (2h-2 (\Delta_{\phi} +1))^2 \Gamma \left(\Delta_{\phi} +\frac{5}{2}\right)}{\sqrt{\pi } 2h (2h+2) (2h-4 \Delta_{\phi} -6) (2h-2 \Delta_{\phi} -3) (2h-4 (\Delta_{\phi} +1)) \Gamma^3 (\Delta_{\phi} +2)}\frac{\gamma^{(1)}_{2,0}}{N^2}+\lambda=0
\end{split}
\end{equation}
Again replacing $\lambda$ using equation(\ref{fixc}) and  $\gamma^{(1)}_{1,0}$ from equation (\ref{g1}) we find that,
\begin{equation}
\gamma^{(1)}_{2,0}=\frac{\Delta_\phi  (\Delta_\phi +1) h (h+1) (-2 \Delta_\phi +h-1) (h-2 \Delta_\phi ) (-2 \Delta_\phi +2 h-3)}{4 (4 \Delta_\phi  (\Delta_\phi +2)+3) (-\Delta_\phi +h-2) (\Delta_\phi -h+1)^2} \gamma^{(1)}_{0,0}
\end{equation}
We can find that in general we will get,
\begin{equation} \label{treeanm}
\gamma^{(1)}_{n,0}=\frac{\Gamma (2 \Delta_\phi ) \Gamma (-h+\Delta_\phi +1)^2 \Gamma (h+n) \Gamma (n+\Delta_\phi )^2 \Gamma (-h+n+2 \Delta_\phi ) \Gamma (-2 h+2 n+2 \Delta_\phi +1)}{n! \Gamma (\Delta_\phi )^2 \Gamma (h) \Gamma (2 \Delta_\phi -h) \Gamma (2 (n+\Delta_\phi )) \Gamma (-h+n+\Delta_\phi +1)^2 \Gamma (-2 h+n+2 \Delta_\phi +1)} \gamma^{(1)}_{0,0}.
\end{equation}
Since we have added one contact term with one undetermined constant, so we loose one equation to fix it which in turn fix all other anomalous dimension in terms of $\gamma^{1}_{0,0}$ as it was found in \cite{kss}. Also notice that the contact term is only present for $\ell'=0$, so we can look for $\ell'=2$ and $s=\Delta_{\phi}$ equation, where $s-$channel will give us,
\begin{equation}
\frac{4^{\Delta_{\phi}} \Gamma(\frac{5}{2}+\Delta_{\phi})\gamma^{(1)}_{0,2}}{\sqrt{\pi}\Delta_{\phi}(1+\Delta_{\phi})\Gamma^3(\Delta_{\phi})}
\end{equation}

whereas the $t-$channel contribution will again start from $O(1/N^4)$ and there is no contribution from contact term, unlike the $\ell'=0$ case, so the solution for anomalous dimension, $\gamma_{0,2}$, is zero and this is true for all non-zero spins. Hence for a theory with $\phi^4$ interaction in the bulk there is no corrections  to anomalous dimensions of spinning double trace operators at $O(1/N^2)$.

\subsection{Exchange}
  Now we outline the procedure to find anomalous dimension of double field operators at $O(1/N^2)$ due to the exchange of a singlet in $t$-channel. This is a different situation from previous case where there was no exchange of any operators other than double fields. Here apart from identity and double field operators we also have exchange of a singlet which appears in the leading order itself. We are considering it here as this would be necessary to reconstruct the triangle diagram we consider in section \ref{triangle}. Let's consider the case when there is a exchange of $\phi$ in crossed channel since this would be relevant for fixing residues of triangle diagram. It was shown in \cite{PDKGAS} that in this situations the ambiguities won't contribute, therefore we can use (\ref{mellineq}) to find $\gamma^{(1)}_{n,\ell}$. We just give an example at $s=\Delta_{\phi}$. 
We choose $\ell'=0$, then only $\ell=0$ and $\Delta_{0,0}=2\Delta_{\phi}+\frac{\gamma^1_{0,0}}{N^2}$ operator contributes in $s$-channel to $O(1/N^2)$. The equation becomes,
\begin{equation}
\frac{\gamma^{(1)}_{0,0} 2^{2 \Delta _{\phi }-1} \Gamma \left(\Delta _{\phi }+\frac{1}{2}\right)}{\sqrt{\pi } \Gamma^3 \left(\Delta _{\phi }\right)}.
\end{equation}
For $\ell=0$ there is also a contribution for the operator $\phi$ itself in $s$- channel,
\begin{equation}
-\frac{C_{\Delta_{\phi},0}  \Gamma \left(\Delta _{\phi }\right) \Gamma \left(-h+\Delta _{\phi }+1\right) \Gamma \left(2 \Delta _{\phi }-h\right)}{\Gamma \left(\frac{\Delta _{\phi }}{2}+1\right) \Gamma^5 \left(\frac{\Delta _{\phi }}{2}\right) \Gamma \left(\frac{3 \Delta _{\phi }}{2}-h\right) \Gamma \left(-h+\frac{3 \Delta _{\phi }}{2}+1\right)}.
\end{equation}
The crossed channel equation for $\phi$ exchange becomes,
\begin{equation}
\begin{split}
& \frac{C_{\Delta_{\phi},0}  \left(\Delta _{\phi }-1\right) \Gamma \left(\Delta _{\phi }-1\right) \Gamma \left(2 \Delta _{\phi }\right) \Gamma \left(\Delta _{\phi }+1\right) \left(\Delta _{\phi }-h\right) \Gamma \left(\Delta _{\phi }-h\right) \Gamma^2 \left(\frac{3 \Delta _{\phi }}{2}-h\right)}{\Gamma \left(\frac{\Delta _{\phi }}{2}+1\right) \Gamma^4 \left(\frac{\Delta _{\phi }}{2}\right)\Gamma^2 \left(\frac{1}{2} \left(3 \Delta _{\phi }-2 h\right)\right)}\\
& \, _5\tilde{F}_4\left(\Delta _{\phi },1-\frac{\Delta _{\phi }}{2},\frac{\Delta _{\phi }}{2},1-\frac{\Delta _{\phi }}{2},h;\frac{3 \Delta _{\phi }}{2},\frac{\Delta _{\phi }}{2}+1,\frac{3 \Delta _{\phi }}{2},-h+\Delta _{\phi }+1;1\right).
\end{split}
\end{equation}
Now using equation (\ref{mellineq}) and solving for $\gamma^{(1)}_{0,0}$ we find,
\begin{equation} \label{exchanm}
\begin{split}
\gamma^{(1)}_{0,0}=& C_{\Delta_{\phi},0}\frac{ 2^{1-2 \Delta _{\phi }} \sqrt{\pi }  \Gamma^3 \left(\Delta _{\phi }\right) \, _5\tilde{F}_4\left(\Delta _{\phi },1-\frac{\Delta _{\phi }}{2},\frac{\Delta _{\phi }}{2},1-\frac{\Delta _{\phi }}{2},h;\frac{3 \Delta _{\phi }}{2},\frac{\Delta _{\phi }}{2}+1,\frac{3 \Delta _{\phi }}{2},-h+\Delta _{\phi }+1;1\right) }{\Gamma^5 \left(\frac{\Delta _{\phi }}{2}\right) \Gamma \left(\Delta _{\phi }+\frac{1}{2}\right) \Gamma \left(\frac{ \Delta _{\phi }+2}{2}\right) \Gamma \left(\frac{3 \Delta _{\phi }}{2}-h\right)}\\
& \bigg(\frac{\Gamma \left(\Delta _{\phi }\right) \Gamma \left(-h+\Delta _{\phi }+1\right) \Gamma \left(2 \Delta _{\phi }-h\right)}{\Gamma \left(-h+\frac{3 \Delta _{\phi }}{2}+1\right)}-2 \left(\Delta _{\phi }-1\right) \Gamma \left(\Delta _{\phi }-1\right) \Gamma \left(\frac{\Delta _{\phi }}{2}\right) \Gamma \left(2 \Delta _{\phi }\right) \Gamma \left(\Delta _{\phi }+1\right)  \\
& \Gamma \left(\Delta _{\phi }-h+1\right) \Gamma \big(\frac{3 \Delta _{\phi }}{2}-h\big) \bigg).
\end{split}
\end{equation}

\begin{figure}[hbt] \label{plot}
	\begin{tabular} {c c c}	
			\includegraphics[width=0.31\textwidth,height=3.5cm]{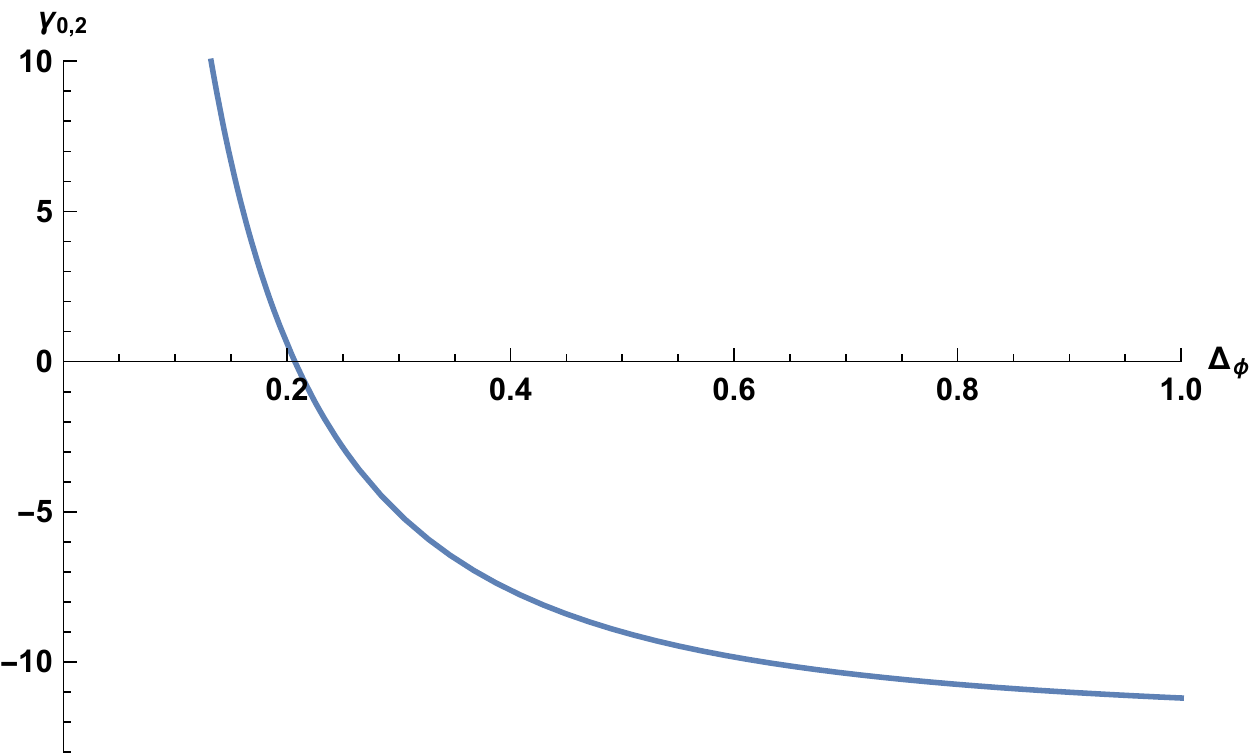} &\includegraphics[width=0.31\textwidth, height=3.5cm]{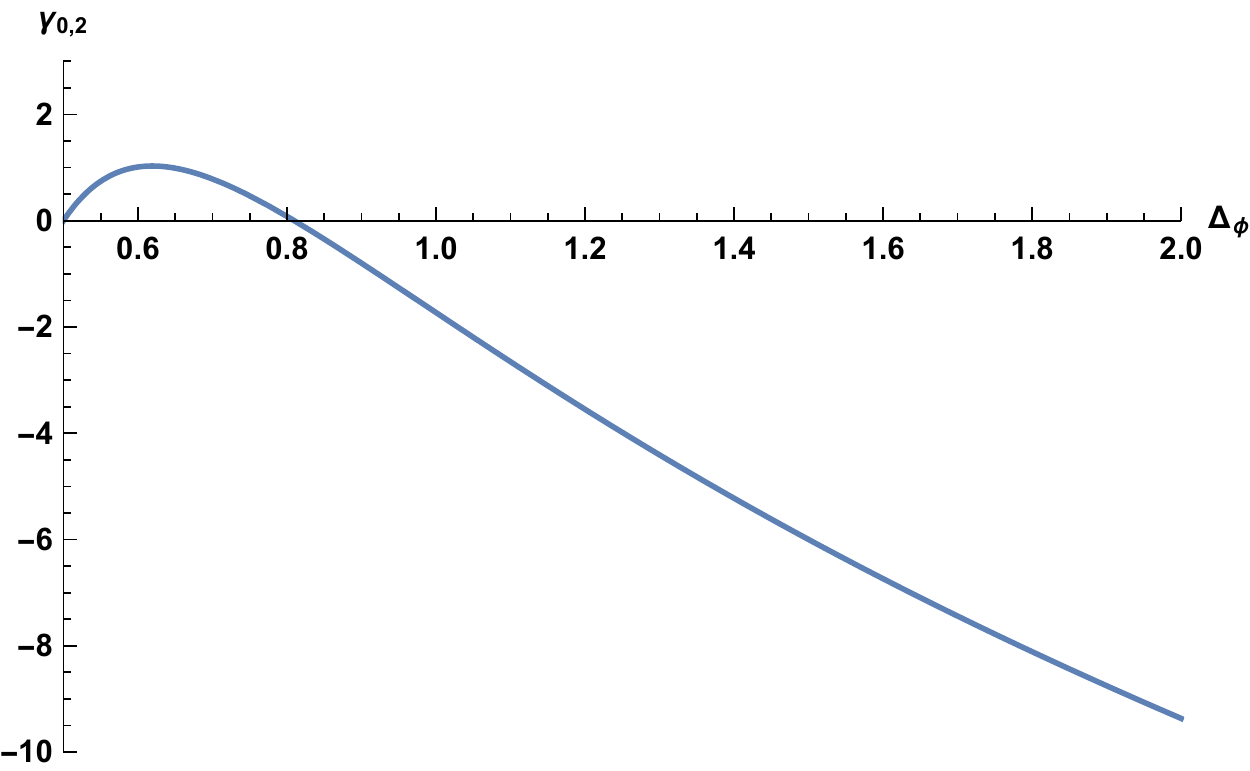} & \includegraphics[width=0.31\textwidth,height=3.5cm]{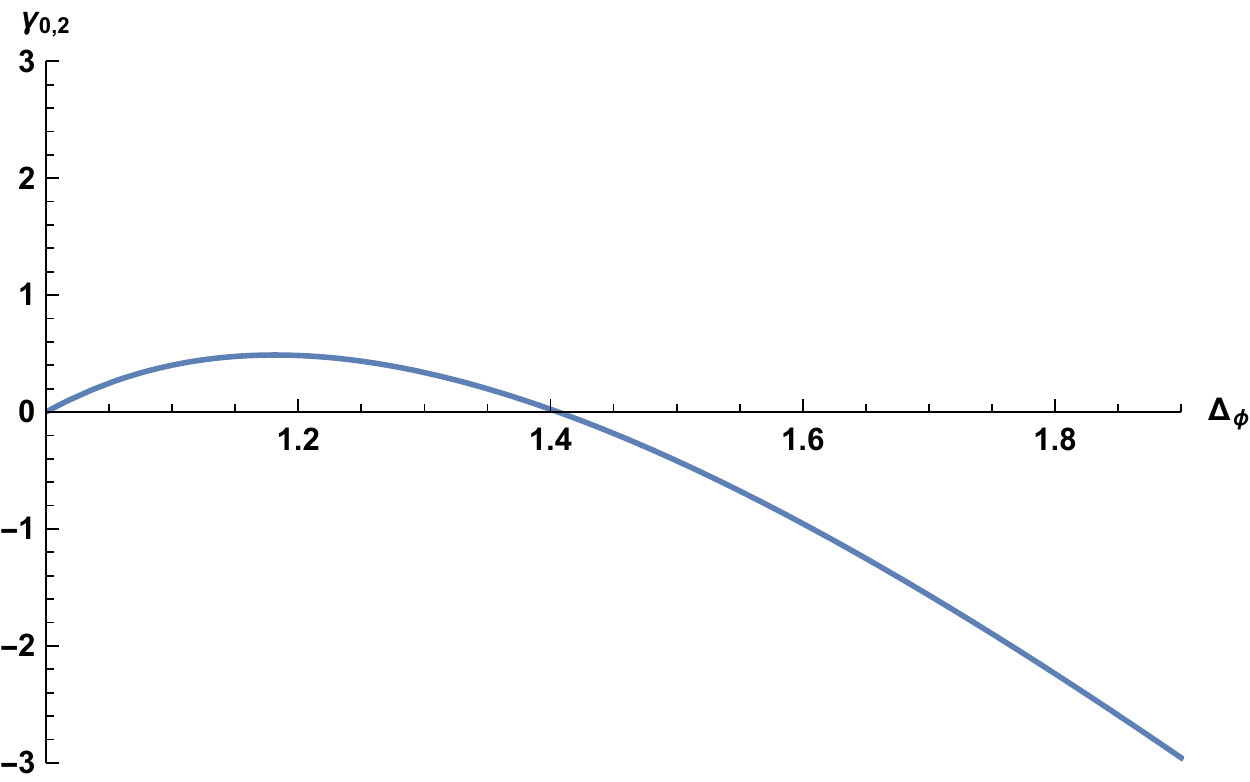}\\
		$ d=2$ & $d=3$ & $d=4$ \\
	\end{tabular} 
	\caption{Plot of $\gamma^{(1)}_{0,2}$ due to stress tensor exchange in different space-time dimensions.}
\label{plot}
\end{figure}

\begin{figure}[hbt]
\centering
\includegraphics[width=0.5\textwidth,height=3.5cm]{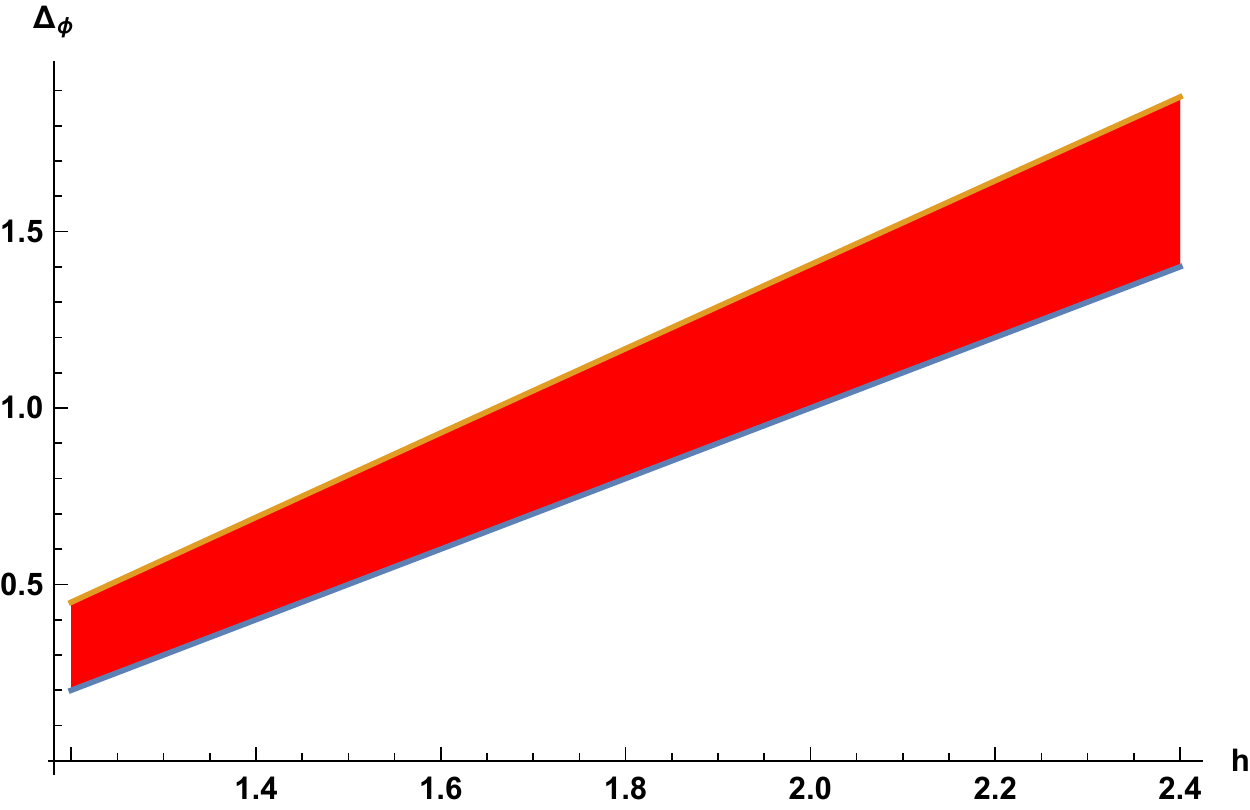}
\caption{Exclusion Plot where $\Delta_{\phi}$ is plotted along the vertical axis and horizontal axis is $h=\frac{d}{2}$. The red region lies above the unitarity bound but it will give positive anomalous dimensions and therefore it might not give us a consistent bulk theory.}
\label{figure2}
\end{figure}

This process of extracting anomalous dimension works similarly at other values of s. One should be careful to take into account all the operators contributing for a specific value of $\ell$. This can be implemented in Mathematica very easily, but unlike $\phi^4$ theory it's not possible to always come up with a closed form formula for $\gamma^1_{n,0}$ for general value of $\Delta_{\phi}$. It was found in four dimensions \cite{adsexchange} \footnote{Also see \cite{Sleight:2018ryu} for derivation of anomalous dimension from crossing kernel in Mellin space.} that the anomalous dimension of spin 2 becomes positive When there is only stress tensor exchange.  From the point of view of AdS/CFT the sign of anomalous dimension is identified with the sign of gravitational force. Now it can be seen from the plots Figure \ref{plot} that $\gamma_{0,2}$ is positive in a regime of unitary values of $\Delta_{\phi}$. This is observed in all dimensions plotted above (figure \ref{figure2}) . Now what does this positivity means in a legit theory of gravity is an interesting question to explore in future.

\subsection{Analysis at $O(1/N^4)$} \label{N4results}

In this section we will show how to extract the CFT data at $O(1/N^4)$ for $\phi^4$ theory solving the constraint equations coming from Polyakov Mellin approach. Unlike $O(1/N^2)$ at this order the solution will have infinite support in spin, which means $\gamma^2_{n,\ell}$ is non-zero for all spins. First we evaluate anomalous dimensions at large spin limit using Witten diagram. So $\gamma^{(2)}_{0,\ell}$ will be given as,
\begin{equation} \label{loopgammaexp}
\gamma^{(2)}_{0,\ell}=\sum_{n=0}^{\infty}\frac{\gamma^{(2) }_n}{J^{2\Delta_{\phi}+2n}}.
\end{equation}															
We will give  closed form general expression for $\gamma^{(2) }_n$ in any space-time dimension and for external scalar operators of any dimension $\Delta_{\phi}$. To do that we first note that for non-zero spins we have,
\begin{equation}
\Delta_{n,\ell}=2\Delta_{\phi}+2n+\ell+\frac{\gamma^{(2)}_{n,\ell}}{N^4}.
\end{equation}
We substitute this in $s-$channel and expand it to find,
\begin{equation}
\frac{ (-1)^{\ell } \left((-1)^{\ell }+1\right) 2^{2 \Delta _{\phi }+\ell -2} \Gamma \left(\ell +\Delta _{\phi }+\frac{1}{2}\right)}{\sqrt{\pi } \Gamma (\ell +1) \Gamma \left(\Delta _{\phi }\right){}^2 \Gamma \left(\ell +\Delta _{\phi }\right)}\frac{\gamma^{2}_{0,\ell}}{N^4}.
\end{equation}

The $t-$channel is given by,
\begin{equation}
W^{t}_{\Delta,\ell}(s,t)=\sum_{q=0}^{\infty}\frac{2 \Gamma \left(-h+\frac{\ell }{2}+\frac{\Delta }{2}+\Delta _{\phi }\right){}^2 \Gamma \left(q+\frac{\Delta }{2}-\Delta _{\phi }-\frac{\ell }{2}+1\right){}^2 f_{p}^{\ell}\left(\Delta _{\phi }+t,s-\Delta _{\phi }\right)}{\Gamma (q+1) \Gamma (-h+q+\Delta +1) \Gamma \left(\frac{1}{2} \left(-\ell +\Delta -2 \Delta _{\phi }+2\right)\right){}^2 \left(-2 \Delta _{\phi }+\Delta +2 q-2 t-\ell \right)}.
\end{equation}
In continuous Hahn basis this becomes,
\begin{equation}
\begin{split}
q^{t}_{\Delta,\ell'|\ell}(s)&=(\kappa_{\ell'}(s))^{-1}\int [dt] \Gamma^2(t+\Delta_{\phi}) \Gamma^2(-t) W^{t}_{\Delta,\ell}(s,t)Q^{2s+\ell'}_{\ell',0}(t)\\
& =(\kappa_{\ell'})^{-1}\int [dt] \Gamma^2(t+\Delta_{\phi}) \Gamma^2(-t) W^{t}_{\Delta,\ell}(s,t) \frac{2^{\ell' } \left((s)_{\ell' }\right){}^2 \, _3F_2(-\ell' ,2 s+\ell' -1,s+t;s,s;1)}{(2 s+\ell' -1)_{\ell' }}.
\end{split}
\end{equation}
So our bootstrap equation is,
\begin{equation}
\sum_{\Delta,\ell}c_{\Delta,\ell}q^{s}_{\Delta,\ell'|\ell}(s)|_{s=\Delta_{\phi}+r}=-2\sum_{\Delta,\ell}c_{\Delta,\ell}q^{t}_{\Delta,\ell'|\ell}(s)|_{s=\Delta_{\phi}+r}.
\end{equation}
We evaluate the bootstrap equation at $s=\Delta_{\phi}$ and also we multiply both side of the equations by,
\begin{equation}
\beta_{\ell}=\kappa_{\ell}(\Delta_{\phi})\bigg(\frac{2^{\ell } \left((s)_{\ell }\right){}^2}{(2 s+\ell -1)_{\ell }}\bigg)^{-1}.
\end{equation}
So our final equation for the anomalous dimension due to  exchange of scalar operator with dimension $\Delta$ has the following form,
\begin{equation} \label{finalg2}
\gamma^{(2)}_{0,\ell}=c_{\Delta,s}\int [dt] \Gamma^2(t+\Delta_{\phi}) \Gamma^2(-t)W^{t}_{\Delta,0}(s,t)\, _3F_2(-\ell ,2 s+\ell -1,s+t;s,s;1).
\end{equation}
At this stage we want to take the large spin limit of ${}_3F_2$ in the right hand side using (\ref{Qasym}) and putting (\ref{loopgammaexp}) on the left side comparing power of J  we find,

\begin{equation}
\begin{split}
\gamma^{(2)}_m =& \sum_{i}\frac{1}{2}(\gamma^{1}_{n,0})^2c_{2\Delta_{\phi}+2n,0}(-\frac{\Gamma (n+q+1)^2 \Gamma \left(-h+n+2 \Delta _{\phi }\right){}^2}{\Gamma (n+1)^2 \Gamma (q+1) \Gamma \left(-h+2 n+q+2 \Delta _{\phi }+1\right)})\Gamma^2(\Delta_{\phi})\Gamma^2(1+n+r+q)\\
&  (\Delta_{\phi}+q+n)_r d(\alpha_1,-\alpha_1,k_1)d(\alpha_2,-\alpha_2,k_2)J^{-2(k_1+k_2+n+r+q+\Delta_{\phi})},
\end{split}
\end{equation}
where $i=k_1+k_2+n+r+q=m$. This gives us closed form expression for coefficients appearing in large spin expansion of $\gamma^2_{0,\ell}$ in any space-time dimensions.\\

 We write first few terms of (\ref{loopgammaexp}) explicitly below,
\begin{equation}
\begin{split}
& \gamma^{(2)}_0=-\Gamma (2 \Delta_{\phi} ),\\
& \gamma^{(2)}_1=-\frac{\Delta _{\phi } \Gamma \left(2 \Delta _{\phi }\right) \left(\Delta _{\phi } \left(\Delta _{\phi } \left(\Delta _{\phi } \left(2 \Delta _{\phi }+4 h-3\right)+h (2-3 h)-7\right)+5 h-3\right)+h-1\right)}{3 \left(2 \Delta _{\phi }+1\right) \left(\Delta _{\phi }-h+1\right)},\\
\end{split}
\end{equation}
\begin{equation}
\begin{split}
& \gamma^{(2)}_2=\frac{\Gamma \left(2 \Delta _{\phi }+3\right)}{720 \left(2 \Delta _{\phi }+1\right){}^2 \left(2 \Delta _{\phi }+3\right) \left(-\Delta _{\phi }+h-2\right) \left(\Delta _{\phi }-h+1\right){}^2}\\
& \bigg(\Delta _{\phi } \big(\Delta _{\phi } \big(\Delta _{\phi } \big(\Delta _{\phi } \big(2 \Delta _{\phi } \big(\Delta _{\phi } \big(\Delta _{\phi } \big(4 \Delta _{\phi } \big(5 \Delta _{\phi }+15 h+4\big)+4 h (53-15 h)-225\big)+h (2 (109-70 h) h\\
&-371)-300\big)+h (h (h (165 h-776)+2327)-2071)+795\big)+h (3 h (h (5 (65-6 h) h-1524)+2873)\\
& -8080)+4326\big)-2 h (h (2 h (45 (h-8) h+1238)-3961)+4342)+3860\big)+h (h (h (45 (19-2 h) h\\
&-2198)+4697)-4798) +1570\big)-18 (h-1) (h (35 h-51)+20)\big)-36 (h-2) (h-1)^2\bigg).
\end{split}
\end{equation}
In $d=2$,
\begin{equation}
\gamma^{(2)}_{0,\ell}=-\frac{2}{J^3}+\frac{1}{4J^5}-\frac{3}{64 J^7}+\frac{5}{512 J^9}-\frac{35}{16384 J^{11}}+.....\,\,\, \text{for}\,\,\Delta_{\phi}=\frac{3}{2}.
\end{equation}
In $d=4$,
\begin{equation}
\gamma^{(2)}_{0,\ell}=-\frac{2}{J^3}-\frac{19}{8 J^5}-\frac{189}{32 J^7}-\frac{11393}{1024 J^9}-\frac{374801}{16384 J^{11}}+...\,\,\,\,\text{for}\,\, \Delta_{\phi}=\frac{3}{2}.
\end{equation}
This will enable us to reconstruct the four point one loop amplitude in $\phi^4$ theory completely.We could have also evaluated the expression (\ref{finalg2}) exactly and that will give us the following form for anomalous dimension,
\begin{equation}
\gamma^{(2)}_{0,\ell}=\sum_{n}\gamma^{(2)}|_{n,0},
\end{equation}
where,
\begin{equation} \label{anmn0}
\begin{split}
\gamma^{(2)}|_{n,0}=&(-1)^{\ell} 2^{-1-2h+4\Delta_{\phi}}  (h-2(n+\Delta_{\phi}))\Gamma(\ell+1) \Gamma(h+n) \Gamma^2(\frac{1}{2}+\Delta_{\phi}) \Gamma^2(\ell+\Delta_{\phi}) \Gamma^2(1-h+\Delta_{\phi}) \\ 
&\frac{\Gamma^3(n+\Delta_{\phi}) \Gamma(\frac{\ell}{2}+n+\Delta_{\phi})\Gamma^3(-h+n+2\Delta_{\phi}) \Gamma(1+\ell+2n+2\Delta_{\phi})\Gamma(\frac{1}{2}-h+n+\Delta_{\phi})}{\pi\Gamma(h)\Gamma(1+n)\Gamma^2(\Delta_{\phi})\Gamma(\frac{1}{2}+n+\Delta_{\phi})\Gamma(1-h+n+\Delta_{\phi})\Gamma^2(-h+2\Delta_{\phi})\Gamma(1-2h+n+2\Delta_{\phi})}\\
& \tilde{W}(a,\ell,\Delta_{\phi},n,h) (\gamma^{(1)}_{0,0})^2,
\end{split}
\end{equation}
where $\tilde{W}$ is the regularized hypergeometric function and W is given by,
\begin{align}
 & W(a,\ell,\Delta_{\phi},n,h)\nonumber\\
& =\,_7F_6 \left(\begin{matrix} \ell+2a, & 1+\frac{\ell}{2}+a, & 1+n, & 1+n, & a, & a, & \ell+h \\ \frac{\ell}{2}+a, & \ell+a+\Delta_{\phi}, & 1+\ell+a, & 1+\ell+a, & 1+a+\Delta_{\phi}, & 1-h+2 a 
\end{matrix} ;1\right),
\end{align}
and $a=n+\Delta_{\phi}$.\\
To compare it with known result we evaluate it for few cases when $\Delta_{\phi}=2$ and $d=4$,
\begin{equation}
\begin{split}
 &\hat{\gamma}^{(2)}_{0,2}|_{0,0}=-\frac{260}{9}+24 \zeta (3), \,\,\,\,\,\,\,\,\,\,\,\,\,\,\,\,\,\,\,\,\,\,\,\,\,\,\,\,\, \hat{\gamma}^{(2)}_{0,4}|_{0,0}=-\frac{259703}{9000}+24 \zeta (3),\\
& \hat{\gamma}^{(2)}_{0,2}|_{1,0}=-\frac{315728}{45}+\frac{29184 \zeta (3)}{5}, \,\,\,\,\,\,\,\,\,\,\,\,\hat{\gamma}^{(2)}_{0,4}|_{1,0}=-\frac{95549104}{5625}+\frac{70656 \zeta (3)}{5},\\
& \hat{\gamma}^{(2)}_{0,2}|_{2,0}=-\frac{9400941}{35}+\frac{7820712 \zeta (3)}{35},\,\,\,\,\hat{ \gamma}^{(2)}_{0,4}|_{2,0}=-\frac{51799815009}{35000}+\frac{43092648 \zeta (3)}{35}.
\end{split}
\end{equation}

These agree with results found in \cite{loopal}\footnote{$\hat{\gamma}^{2}_{0,\ell}|_{n,0}=\frac{\gamma^{(2)}_{0,\ell}|_{n,0}}{(\gamma^{(1)}_{0,0})^2} $. Also $\gamma^{(2)}_{0,\ell}|_{n,0}$ is that of \cite{loopal} multiplied by $\frac{1}{8} C_{n,0}(\gamma^{(1)}_{n,0})^2$. }. Here we give a closed form formula \ref{anmn0} valid for any general case.

Now we turn our attention to a case where $\Delta_\phi=2$ in 3 dimensions and we perform the explicit sums in crossed channel to give a closed form answer for CFT data down to spin zero. We compare this with recently found answer in \cite{loop4} . This provides an explicit evidence where we produce the loop results using Polyakov Mellin bootstrap. Since we are bootstrapping $\phi^4$ theory in the bulk therefore the contact term we add in mellin space is a constant. So it will only contribute to $\ell'=0$ equation. So we can look at $s=\Delta_{\phi}$ and $\ell'=2$ equation where we don't have a contact term therefore our equation is simply,
\begin{equation} \label{spin2eq}
\sum_{\Delta,\ell}c_{\Delta,\ell} (q^{s}_{\Delta,2|\ell}(\Delta_{\phi})+2q^{t}_{\Delta,2|\ell}(\Delta_{\phi})=0
\end{equation} 

In the $s-$channel the only operator $n=0$ with spin 2 contributes till $O(1/N^4)$,

\begin{equation}
\sum_{\Delta,\ell}c_{\Delta,\ell} q^{s}_{\Delta,\ell'|\ell}=\frac{35}{2}\gamma_{0,2}
\end{equation}
In the $t-$ channel only the scalars with dimensions $\Delta=4+2n$ contribute,
\begin{equation}
\begin{split}
\sum_{\Delta,\ell}c_{\Delta,\ell} q^{t}_{\Delta,\ell'|\ell}=&\sum_{n}\frac{2 (4 n+5) \gamma_{n,0}^2}{\sqrt{\pi }} \Big[\frac{35}{144} \sqrt{\pi } \left(16 (n+1)^6+40 (n+1)^5+104 (n+1)^4+146 (n+1)^3\right.\\
&\left. +133 (n+1)^2+77 n+98\right)+ \frac{35}{72} \sqrt{\pi } (n+1)^2 (2 n+3)^2 (n (2 n+5) (n (2 n+5)+14)\\
&+42)  \left(\psi ^{(1)}\left(n+\frac{3}{2}\right)-\psi ^{(1)}(n+1)\right)\Big]
\end{split}
\end{equation}
Now solving our bootstrap equation (\ref{spin2eq}) we find,
\begin{equation} \label{spin2n0}
\gamma^{(2)}_{0,2}=-\frac{1}{20} (\gamma^{(1)}_{0,0})^2.
\end{equation}
Now we can see the constraint equations coming from vanishing of residue of double  pole contribution at $s=\Delta_{\phi}+1$. and $\ell'=0$.  Now we will have a contribution coming from $\phi^4$ contact term, therefore our constraint equation is,
\begin{equation}\label{sdphi1}
\sum_{\Delta,\ell} c_{\Delta,\ell}\Big(q^{s}_{\Delta,0|\ell}(\Delta_{\phi}+1)+2 q^{t}_{\Delta,0|\ell}(\Delta_{\phi}+1)\Big)+\lambda=0
\end{equation}
As explained before we can fix $\lambda$ by loosing one equation, here we choose the constraint equation coming from vanishing of residue of double pole at $s=\Delta_{\phi}$ for $\ell'=0$,
\begin{equation}
\sum_{\Delta,\ell} c_{\Delta,\ell}\Big(q^{s}_{\Delta,0|\ell}(\Delta_{\phi})+2 q^{t}_{\Delta,0|\ell}(\Delta_{\phi})\Big)+\lambda=0
\end{equation}
Therefore we have,
\begin{equation}
\lambda=-\sum_{\Delta,\ell} c_{\Delta,\ell}\Big(q^{s}_{\Delta,0|\ell}(\Delta_{\phi})+2 q^{t}_{\Delta,0|\ell}(\Delta_{\phi})\Big)
\end{equation}
Replacing $\lambda$ in equation (\ref{sdphi1}) we get,
\begin{equation}
\sum_{\Delta,\ell} c_{\Delta,\ell}\Big(q^{s}_{\Delta,0|\ell}(\Delta_{\phi}+1)+2 q^{t}_{\Delta,0|\ell}(\Delta_{\phi}+1)\Big)-\sum_{\Delta,\ell} c_{\Delta,\ell}\Big(q^{s}_{\Delta,0|\ell}(\Delta_{\phi})+2 q^{t}_{\Delta,0|\ell}(\Delta_{\phi})\Big)=0
\end{equation}
In the $s-$channel spin zero and spin 2 contributes. For spin zero the operators with dimension $\Delta_{0,0}$ and $\Delta_{1,0}$ and for spin 2, the operator with dimension $\Delta_{0,2}$ contribute. In the crossed channel only spin zero $\Delta_{n,0}$ contributes. Using $\gamma_{0,2}$ found in equation (\ref{spin2n0}) we find,
\begin{equation}
(\gamma^{(2)}_{1,0}-\gamma^{(2)}_{0,0})=\frac{7}{5} (\gamma^{(1)}_{0,0})^2.
\end{equation}
Similarly we can find all data (i.e. for other values of $n$ and spins) at this order by looking at constraint equations coming from other poles $s=\Delta_{\phi}+n$ and the basis spin $\ell'$.

\section{Loops in AdS}

In this section we briefly review how to compute loops in AdS from CFT data following \cite{loopal}. We consider the $1/N$ expansion of connected part of four point function $g(u,v)$ and the same for Mellin amplitude $M(s,t)$,
\begin{equation} \label{largenexp}
\begin{split}
& g(u,v)=\frac{g^{(1)}(u,v)}{N^2}+\frac{g^{(2)}(u,v)}{N^4}+....\\
& M(s,t)=\frac{M^{(1)}(s,t)}{N^2}+\frac{M^{(2)}(s,t)}{N^4}+....
\end{split}
\end{equation}
We know from (\ref{sch}) that,
\begin{equation}\label{sch1}
\begin{split}
g(u,v)&=\sum_{\Delta,\ell}C_{\Delta,\ell}g_{\Delta,\ell}(u,v)\\
& = \sum_{n,\ell}C_{\Delta_{n,\ell},\ell}u^{\frac{\tau_{n,\ell}}{2}}\tilde{g}_{\Delta_{n,\ell}}(u,v)
\end{split}
\end{equation}
Plugging (\ref{correction}) in (\ref{sch1}) and expanding in $1/N$ we find,
\begin{equation}
g^{(1)}(u,v)=\sum_{n,\ell} u^{\Delta_{\phi}+n}\left(C_{n,\ell}^1+\frac{1}{2}C_{n,\ell}^0\gamma_{n,\ell}^{(1)}\left(\log u+\frac{\partial}{\partial n}\right)\right)\tilde{g}_{2\Delta_{\phi}+2n+\ell,\ell}.
\end{equation}
At order $1/N^4$ the correlator becomes,
\begin{equation}
\begin{split}
g^{(2)}(u,v)=&\sum_{n,\ell} u^{\Delta_{\phi}+n}\bigg(C_{n,\ell}^2 +\frac{1}{2}C_{n,\ell}^{0}\gamma_{n,\ell}^{(2)}(\log u +\frac{\partial}{\partial n})\\
& \frac{1}{2}C_{n,\ell}^{(1)} \gamma_{n,\ell}^1(\log u+\frac{\partial}{\partial n})\\
& +\frac{1}{8}C_{n,\ell}^0(\gamma_{n,\ell}^{(1)})^2\left(\log^2( u) +2 \log u \frac{\partial}{\partial n}+\frac{\partial^2}{\partial n^2}\right)\bigg)\tilde{g}_{2\Delta_{\phi}+2n+\ell,\ell}.
\end{split}
\end{equation}
Now we focus on $u^{\Delta_{\phi}}\log u$ term that has infinite support in spin and it's Mellin transform,
\begin{equation}\label{N4exp}
\begin{split}
&\sum_{\ell}C_{0,\ell}^{0}\frac{\gamma_{0,\ell}^{(2)}}{(2)}\tilde{g}^0_{2\Delta_{\phi}+\ell}(v)\\
& =\sum_{\ell}\gamma_{0,\ell}^{(2)}\frac{\Gamma(2\Delta_{\phi}+\ell-1)}{\ell!\Gamma^4(\Delta_{\phi})}(2\Delta_{\phi}+2\ell-1)\int_{-i \infty}^{i \infty} [dt]v^t \Gamma^2(t+\Delta_{\phi})\Gamma^2(-t)\\
& \times {}_3F_2\bigg[\begin{matrix} -\ell,\, 2\Delta_{\phi}+\ell-1,\,\Delta_{\phi}+t\\
\ \ \Delta_{\phi} \ \ , \  \  \Delta_{\phi}
\end{matrix};1\bigg] .
\end{split}
\end{equation}
The highest $\log$ term appeared in (\ref{N4exp}) is $\log^2(u)$. It can be seen in (\ref{mellin}) that one can only get a $\log(u)$ term from the double pole in s in the measure. But we can plug in expansion of $M(s,t)$ as given in (\ref{largenexp}) and $M^{(2)}(s,t)$ should reproduce (\ref{N4exp}).  So to reproduce $\log^2(u)$ we need $M^{(2)}(s,t)$ to have the following structure,
\begin{equation}
M^{(2)}(s,t)=\sum_{n}\frac{R_n}{s-\Delta_{\phi}-n}+crossing.
\end{equation}
Then it is clear that from measure in (\ref{mellin}) we have double poles at $s=\Delta_{\phi}+n$ and then because of these proposed poles above  in $M^{(2)}(s,t)$ we will have altogether triple poles in $s$ at double field locations which will reproduce the $u^{\Delta_{\phi}+n}\log^2(u)$ terms. Besides these we can always add a crossing symmetric regular function of $s$ and $t$ to this amplitude. But we can't fix them using PM bootstrap, so we only focus on the pole piece of $M^{(2)}(s,t)$ in this paper. 
Now the term proportional to $u^{\Delta_{\phi}}\log u$ at $O(1/N^4)$ is,
\begin{equation}
\int_{-i\infty}^{i \infty} [dt] \Gamma(\Delta_{\phi}+t)^2 \Gamma(-t)^2 \tilde{M}^{(2)}(s,t).
\end{equation} 
Comparing this with (\ref{N4exp}) and using the orthonormality of continuous Hahn polynomial one finds that\footnote{For $\ell=0$ there will be a contribution from $s=\Delta_{\phi}$ (triple pole) also.},
\begin{equation}\label{loopformula}
\gamma^{(2)}|_{0,\ell>0}= \int_{-i\infty}^{i \infty} [dt]\Gamma^2(\Delta_{\phi}+t) \Gamma^2(-t)\tilde{M}^{(2)}(\Delta_{\phi},t) {}_3F_2\bigg[\begin{matrix} -\ell,\, 2\Delta_{\phi}+\ell-1,\,\Delta_{\phi}+t\\
\ \ \Delta_{\phi} \ \ , \  \  \Delta_{\phi}
\end{matrix};1\bigg] .
\end{equation}
 $ \tilde{M}^{(2)}(\Delta_{\phi},t)$ only includes $t$ and $u$ channel. since  there is a symmetry as  $t\rightarrow -s-t$ so we can work with simply $t$-channel and multiply with factor of 2. Since we have closed form expressions for coefficients appearing for $\gamma^{(2)}_{0,\ell}$ in the large spin limit for general case we can fix the residues using (\ref{loopformula}). We show explicit examples for two theories considered before.

\subsection{$\phi^4$ theory}
\begin{figure}[h]
\begin{tabular}{c}
		\,\,\,\,\,\,\,\,\,\,\,\,\,\,\,\,\,\,\,\,\,\,\,\,\,\,\,\,\,\,\,\,\,\,\,\,\,\,\,\,\,\,\,\,\,\,\,\,\,\,\,\,\,\,\,\,\,\,\,\,\,\,\,\,\,\,\,\,\,\,\,\,\,\,\,\,\,\,\,\,\,\,\,\,\,\includegraphics[width=0.31\textwidth,height=3.5cm]{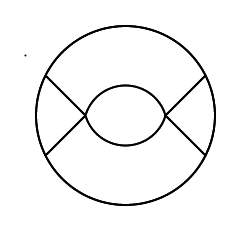}  	
\end{tabular}
	\caption{Bubble diagram in $\phi^4$ theory.}
	\label{figure3}
\end{figure}

First we consider $\phi^4$ theory in $AdS_{d+1}$. So there is only one loop diagram, figure \ref{figure3}. We give results for few values of $\Delta_{\phi}$ in different space-time dimensions.

 We write one loop Mellin amplitude as,
\begin{equation}
M^{(2)}(s,t)=\sum_{n}\frac{R_n}{s-\Delta_{\phi}-n}+crossing
\end{equation}
In $d=4$,
\begin{equation}
\begin{split}
& R_n=-\frac{245 \sqrt{\pi } (n (7 n (11 n (n+8)+239)+1794)+640) \Gamma (n+1)}{12288 \Gamma \left(n+\frac{9}{2}\right)}(\gamma^{(1)}_{0,0})^2,\,\,\,\,\,\, \text{for}\,\, \Delta_{\phi}=4,\\
& R_n=\left(-\frac{8}{\pi ^2 (n+1)}-\frac{8 \Gamma (n+1)}{\pi ^{3/2} \Gamma \left(n+\frac{1}{2}\right)}\right)(\gamma^{(1)}_{0,0})^2,\,\,\,\,\,\,\,\,\,\,\,\,\,\,\,\,\,\,\, \text{for}\,\,\Delta_{\phi}=\frac{3}{2}.
\end{split}
\end{equation} 
In $d=3$,
\begin{equation}
R_n=\left(-\frac{\sqrt{\pi } \Gamma (n+1)}{8 \Gamma \left(n+\frac{3}{2}\right)}-\frac{1}{4}\right)(\gamma^{(1)}_{0,0})^2,\,\,\,\,\,\,\,\text{for}\,\, \Delta_{\phi}=1.
\end{equation}
We can use our method to evaluate residue for any value of mass of the scalar field in $AdS_{d+1}$. But it is not always possible to come up with a simple closed form expression like above easily, though the residues can be systematically fixed by implementing it in Mathematica quite easily. It can be seen that if we try to sum the above residues in one of the channel, say $t$- channel then it will diverge. This UV divergence is expected for $AdS_{d+1\geq 4}$. On the other hand it gives convergent expression for $AdS_{3}$. e.g. in d=2,

\begin{equation}
R_n=-\frac{16 (1)_n}{\pi ^2 \left(\frac{3}{2}\right)_n}(\gamma^{(1)}_{0,0})^2,\,\,\,\, \text{for}\,\, \Delta_{\phi}=\frac{3}{2}.
\end{equation}
If we sum in the t channel this gives,
\begin{equation}
-\frac{16 \, _3F_2\left(1,1,-t;\frac{3}{2},1-t;1\right)}{\pi ^2 t}(\gamma^{(1)}_{0,0})^2
\end{equation}
All the results we quoted here are in agreement with the existing results in the literature \cite{loopal,kaplanloop2,penedonesm}.
\subsection{$\phi^3+\phi^4 $ theory} \label{triangle}
Now we consider a local effective theory in $AdS_{d+1}$ with cubic and quartic both coupling present. So the effective Lagrangian becomes,
\begin{equation}
\mathcal{L}=\frac{1}{2}(\partial\phi)^2+\frac{1}{2}m^2\phi^2+\lambda_3 \phi^3+\lambda_4\phi^4
\end{equation} 

\begin{figure}[hbt]
\centering
\includegraphics[width=0.31\textwidth,height=3.5cm]{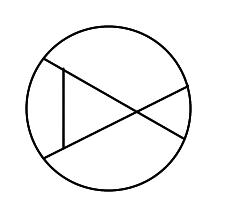}
\caption{Triangle diagram in $\phi^3$+$\phi^4$ theory}
\label{figure4}
\end{figure}
We would consider the triangle diagram which is present in this theory at one loop (figure \ref{figure4}). This is relatively simpler to evaluate using this approach. Because as there is cubic coupling, therefore $\phi$ will appear in the OPE and the solution $\gamma^{(1)}_{n,\ell}|^{\phi^3}$ will have support for all spin. Therefore we need to perform a spin sum also to get $\gamma^{(2)}_{0,\ell}$. But the triangle diagram has both the vertices cubic and quartic in it. Therefore although we have,
\begin{equation}
(\gamma^{(1)}_{n,\ell})^2=(\gamma^{(1)}_{n,\ell}|^{\phi^3})^2+(\gamma^{(1)}_{n,\ell}|^{\phi^4})^2+2\gamma^{(1)}_{n,\ell}|^{\phi^3}\gamma^{(1)}_{n,\ell}|^{\phi^4}
\end{equation}
and $\gamma^{(1)}_{n,\ell}|^{\phi^4}$ is non-zero only for spin zero, we only need the contribution of $2\gamma^{(1)}_{n,0}|^{\phi^3}\gamma^{(1)}_{n,0}|^{\phi^4}$ to $\gamma^{(2)}_{0,\ell}$ for the evaluation of triangle diagram. We already have computed $\gamma^{(1)}_{0,0}|^{\phi^3}$ in (\ref{exchanm}) and $\gamma^{(1)}_{n,0}|^{\phi^4}$ in (\ref{treeanm}). Using these we can fix the loop amplitude completely just the way we did it for bubble diagrams. We can evaluate these for any general case of interest. We have also checked that our results match exactly with the results found in \cite{loopal} which we don't repeat here. As an demonstration we quote  values of few leading residues for two different cases  in Table  \ref{res34}. Indeed for many interesting cases (though mostly limited to $\Delta_{\phi}$ integers or half integers which might have interesting applications when the dimensions are protected) it is possible to find the closed form for residues and perform the sums. To the best of our knowledge this diagram has not been evaluated for general case using any other methods.

\begin{table}[hbt] 
\centering
\begin{tabular}{|l|l|l|l|l|}
\hline
 &  $R_0$ & $R_1$ & $R_2$ & $R_3$ \\ \hline
$d=4$,$\Delta_{\phi}=4$ & $\frac{193}{9}$  & $\frac{1357}{63}$  & $\frac{127937}{6237}$  & $\frac{5789543}{297297} $ \\ \hline
$d=3$,$\Delta_{\phi}=2$ & 5  & 3  & $\frac{20}{9}$  & $\frac{43}{24} $ \\ \hline
\end{tabular}
\caption{Examples of first few residues in $\phi^3+\phi^4$  theory. The numbers shown above need to be multiplied with $\gamma^{(1)}_{0,0} c_s $ to get the final answer. $c_s$ is the OPE coefficient of operator $\phi$, i.e., $C_{\phi \phi \phi}$.}
\label{res34}
\end{table}

\section{Discussion}
In this paper, we have successfully implemented the Polyakov Mellin bootstrap  for holographic conformal field theories following the proposal \cite{pmb}. We calculated all CFT data for holographic scalar theory till one loop and this gives correct result for spin zero as well. Using this data we reconstructed certain one loop diagrams in AdS following the proposal in \cite{loopal}. Our method is particularly useful in odd dimensions where the blocks are not known in closed form. Also we can use our expressions for any general conformal dimension of the external scalar fields. 

 There are several interesting future directions one can pursue:
 
 $\bullet$ For $\phi^3+\phi^4$ theory the total anomalous dimensions at $O(1/N^4)$ will require an infinite spin sum to perform. This was done in [\cite{gravitysmatrix},\cite{csboot}] for specific cases. It will be interesting to see if those sums can be easily done in Mellin space using properties of special functions and if it can be generalized to other cases.

 $\bullet$ It is natural to generalize these techniques for spinning correlator\footnote{See \cite{sleightspin} for discussion on spinning correlator formulation in Mellin space.} which will in turn fix the loop amplitudes involving external spinning fields. Recently this was done for internal scalar exchanges in \cite{6j}.

 $\bullet $ As loops are fixed by tree level data. It will be very pleasing to explore if there is any organizing principles like Feynman rules in AdS \cite{feynmanrule}. Finally to answer questions like What would be the analogues of Optical theorems and generalized unitarity methods of S-matrix in AdS?

 $\bullet$ Finally from the perspective of Polyakov-Mellin bootstrap fixing the ambiguities in the basis completely is the most crucial question. If studying higher order loops can give some insight to it or some physical principle which can fix it that would be a tremendous achievement!\\

\section*{Acknowledgements}
The author would like to thank Aninda Sinha for suggesting the problem and numerous helpful discussions. The author also thanks Apratim Kaviraj, Subham Dutta Chowdhury and Parijat Dey for comments on the draft. 
\appendix

\section{Continuous Hahn Polynomial and its asymptotic}
 The definition of the continuous Hahn polynomial is:

\begin{equation}\label{hahndef}
{Q}^{2s+\ell}_{\ell,0}(t)= \frac{2^{\ell}\,((s)_{\ell})^2}{(2s+\ell-1)_{\ell}}\,{}_3F_2\bigg[\begin{matrix} -\ell,\, 2s+\ell-1,\,s+t\\
	\ \ s \ \ , \ \ \ \ \ \  s
\end{matrix};1\bigg]\,.
\end{equation}
These polynomials satisfy the orthogonality property \cite{AAR},
\begin{equation}\label{orthonorm}
\frac{1}{2\pi i}\int_{-i\infty}^{i\infty} dt \ \G^2(s+t)\G^2(-t) {Q}^{2s+\ell}_{\ell,0}(t) {Q}^{2s+\ell '}_{\ell',0}(t)=(-1)^\ell {\kappa}_{\ell}(s)\d_{\ell,\ell'}\,,
\end{equation}
where,
\begin{equation}\label{kappadefn}
{\kappa}_{\ell}(s)= \frac{4^\ell \ell!}{(2s+\ell-1)_\ell^2}\frac{\G^4(\ell+s)}{(2s+2\ell-1)\G(2s+\ell-1)}\,.
\end{equation}
The large $\ell$ asymptotic of the ${}_3F_2$ appearing in (\ref{hahndef}) was derived in \cite{PDKGAS} and we quote the result here,
\begin{align}\label{Qasym}
{}_3F_2\bigg[\begin{matrix} -\ell,\, 2s+\ell-1,\,s+t\\
\ \ s \ \ , \  \  s
\end{matrix};1\bigg]  & \sim \sum_{n, k_1, k_2=0}^{\infty}\frac{(-1)^n}{n!}\frac{\G^2(s)\,(s+t)_n}{\G(-t-n)^2}\,\mathfrak{b}_{k_1}(s)\,\mathfrak{b}_{k_2, n}(t)\,  J^{-2k_1-2k_2-2n-2s-2t}\nn & 
 +\sum_{n, k_1, k_2=0}^{\infty}\frac{(-1)^n}{n!}\frac{\G^2(s)\,(-t)_n}{\G(s+t-n)^2}\,\mathfrak{b}_{k_1}(s)\,\mathfrak{b}_{k_2, n}(-s-t)\, J^{-2k_1-2k_2-2n+2t}
\end{align}
where,
\be \label{dd3}
J^2=(\ell+s)(\ell+s-1), \qquad \mathfrak{b}_{k_1}(s) = d_{\a_1, \b_1, k_1}, \qquad
 \mathfrak{b}_{k_2, n}(t) = d_{\a_2, \b_2, k_2},
\ee
and
\be
\a_1=1-s=-\b_1, \quad  \a_2=-t-1-n=-\b_2\,. 
\ee
 Also $d_{\a, \b, k}$ is defined as,
\be \label{dd1}
d_{\a, \b, k}= \sum_{j=0}^{k}\, c_j \, \binom{\frac{\a-\b-2j}{2}}{k-j}  \bigg(\frac{-1+\a+\b}{2}\bigg)^{2k-2j}
\ee
and
\be \label{dd2}
c_j = \frac{\G(\b-\a+2j)}{\G(\b-\a)\,(2j)!}\, \mathcal{B}^{1+\a-\b}_{2j} (\frac{1+\a-\b}{2})\,,
\ee
where $\mathcal{B}$ is the generalized Bernoulli polynomial \cite{PDKGAS}.

\section{Contact terms}\label{contact}

In this section we provide another example where we show how Polyakov Mellin bootstrap works for a holographic CFT.  To our basis we add the following contact term,
\begin{equation} \label{2deriv}
u^{\Delta_{\phi}}(1+u+v)\bar{D}_{\Delta_{\phi}+1~\Delta_{\phi}+1~ \Delta_{\phi}+1~ \Delta_{\phi}+1}(u,v)
\end{equation}  
As argued in \cite{pmb} any contact diagram can be parametrized as,
\begin{equation}\label{mellinc}
\sum_{m+n=0}^{\frac{L}{2}}a_{mn} \Big(s(s+t-\Delta_{\phi})(t+\Delta_{\phi})\Big)^m\Big(t(s+t)+s(s-\Delta_{\phi})\Big)^n
\end{equation}
In Mellin space the contact term (\ref{2deriv}) can be written as (\ref{mellinc}) with $a_{01}=2$ and $a_{00}=\Delta_{\phi^2}$. Now we have to decompose it in continuous Hahn basis,
\begin{equation}
c(s,t)=\sum_{\ell'=0}^2a_{\ell'}(s)Q^{2s+\ell'}_{\ell',0}(t)
\end{equation}
with 
\begin{equation}
\begin{split}
a_{0}(s)=& a_{00}-\frac{s}{4s+2}\Big( 2\Delta_{\phi}(a_{10}\Delta_{\phi}+a_{01})+s^3a_{10}+s^2 (a_{10}(1-4\Delta_{\phi})-3a_{01})+s(a_{01}(4\Delta_{\phi}-1)\\
& +2a_{10}\Delta_{\phi}(2\Delta_{\phi}-1))\Big)
\end{split}
\end{equation}

Now our constraint equation becomes,
\begin{equation}
\sum_{\Delta,\ell}c_{\Delta,\ell}(q^{s}_{\Delta,\ell'|\ell}(s)+2q^{t}_{\Delta,\ell'|\ell}(s))+\lambda a_{\ell'}(s) =0
\end{equation}

We look at equation $s=\Delta_{\phi}$ for $\ell'=0$,
\begin{equation}
\frac{\Gamma \left(2 \Delta _{\phi }\right)}{\Gamma^4 \left(\Delta _{\phi }\right)}\gamma_{0,0} +\lambda \frac{\Delta _{\phi }^3}{2 \Delta _{\phi }+1} =0
\end{equation}

 which fix $\lambda$,
\begin{equation} \label{fixamb}
\lambda=-\frac{\left(2 \Delta _{\phi }+1\right) \Gamma \left(2 \Delta _{\phi }\right)}{\Delta _{\phi }^3 \Gamma^4 \left(\Delta _{\phi }\right)} \gamma_{0,0} 
\end{equation} 
Now we can solve for all other quantities in terms of $\gamma_{0,0}$. To see it we take $s=\Delta_{\phi}$ and $\ell'=2$ equation,
\begin{equation}
\frac{ 4^{\Delta_\phi }  \Gamma \left(\Delta_\phi +\frac{5}{2}\right)}{\sqrt{\pi } \Delta_\phi  (\Delta_\phi +1) \Gamma^3 (\Delta_\phi )}\gamma_{0,2}+\frac{\lambda}{2}=0
\end{equation}
Now we replace $\lambda$ using \ref{fixamb} to find,
\begin{equation}
\gamma_{0,2}=\frac{\gamma_{0,0} (\Delta_\phi +1)}{\Delta_\phi ^2 (2 \Delta_\phi +3)}
\end{equation}
Similarly looking at $s=\Delta_{\phi}+1$ and $\ell'=0$ equation we find,
\begin{equation} 
\begin{split}
\gamma_{1,0}=&\frac{\gamma_{0,0}}{\Delta _{\phi }^2 \left(2 \Delta _{\phi }+3\right) \left(\Delta _{\phi }-h+1\right)}\Big(\left(\Delta _{\phi }+1\right){}^2-h^2 \left(\Delta _{\phi } \left(\Delta _{\phi } \left(\Delta _{\phi }+3\right)+5\right)+4\right)\\
& +h \left(\Delta _{\phi } \left(2 \Delta _{\phi } \left(\Delta _{\phi } \left(\Delta _{\phi }+3\right)+6\right)+13\right)+3\right)\Big)
\end{split}
\end{equation}

This agrees with the results found in \cite{kss} in 2 and 4 space-time  dimension. The procedure outlined above can be followed for any contact diagrams in AdS to find the CFT data at $O(1/N^2)$. Then following section \ref{N4results} we can reproduce the loop results.
\section{Useful expressions}
The notations used in \ref{tchexp} are given below,
\begin{equation}
\chi^{(n)}_{\ell'}(s)=(-1)^{\ell'}2^{-\ell'}\frac{\Gamma(2s+2\ell')\Gamma^2(s+n)}{\ell'!\Gamma^2(\ell'+s)\Gamma(2s+n)}\frac{(-n)_{\ell'}}{(2s+n)_{\ell'}}.
\end{equation}
and
\begin{equation}
a=\ell'+2(a_{\ell}+m+s-1),\,\,\,\,\,b=e=a_{\ell}+m,\,\,\,\,c=d=a_{\ell}+m+s-1,\,\,\,\,f=2(s-\Delta_{\phi})+h+m+\ell'-\ell,
\end{equation}
where $a_{\ell}=1+\frac{\Delta-\ell}{2}-\Delta_{\phi}$.
\\The normalization used in PM bootstrap is defined as,
\be\label{normt}
{\mathcal{N}}_{\D, \ell}= \frac{(-2)^{\ell}\,(\ell+\D-1)\,\G(1-h+\D)\,\G^2(\ell+\D-1)}{\G(\D-1)\,\G^4(\frac{\D+\ell}{2})\,\G^2(\frac{\ell+2\D_\f-\D}{2})\,\G^2(\frac{\ell+2 \D_\f+\D-2h}{2})}.
\ee

\end{document}